# A 300-500 MHz Tunable Oscillator Exploiting Ten Overtones in Single Lithium Niobate Resonator

Ali Kourani, Ruochen Lu, Anming Gao, and Songbin Gong
Department of Electrical and Computer Engineering
University of Illinois at Urbana-Champaign
Urbana, IL, USA

*Abstract*—This paper presents the first voltage-controlled MEMS oscillator (VCMO) based on a Lithium Niobate (LiNbO$_3$) lateral overtone bulk acoustic resonator (LOBAR). The VCMO consists of a LOBAR in a closed loop with 2 amplification stages and a varactor-embedded tunable LC tank. By adjusting the bias voltage applied to the varactor, the tank can be tuned to change the closed-loop gain and phase responses of the oscillator so that the Barkhausen conditions are satisfied for a particular resonance mode. The tank is designed to allow the proposed VCMO to lock to any of the ten overtones ranging from 300 to 500 MHz. Owing to the high-quality factors of the LiNbO$_3$ LOBAR, the measured VCMO shows a low close-in phase noise of -100 dBc/Hz at 1 kHz offset from a 300 MHz carrier and a noise floor of -153 dBc/Hz while consuming 9 mW. With further optimization, this VCMO can lead to direct radio frequency (RF) synthesis for ultra-low-power transceivers in multi-mode Internet-of-Things (IoT) nodes.

*Keywords*—lithium niobate, MEMS, oscillators, overtone.

## I. INTRODUCTION

Recently, low-power, low phase noise, wide tuning range and miniature radio frequency (RF) synthesizers are becoming highly desirable for battery-powered multi-mode Internet-of-Things (IoT) transceivers. To this end, voltage-controlled MEMS oscillators (VCMOs) employing high quality factor ($Q$) and high electromechanical coupling ($k_t^2$) acoustic resonators are emerging as a great enabler for such direct RF synthesis. Unfortunately, the tuning range of a VCMO based on a MEMS resonator is ultimately limited by the resonator $k_t^2$ [1-5].

One way to increase the oscillator's tuning range, regardless of the resonator design or its comprising material, is to switch between multi-frequency resonators within an oscillator [6-8]. Although each of these resonators can be solely optimized for $Q$ and $k_t^2$, the system requires more devices to cover a certain RF band and thus leading a larger system prone to fabrication yield issues.

On the other hand, co-designing a VCO with a multi-resonance resonator helps to reduce the parallelism without sacrificing performance. For this reason, dual-mode MEMS oscillators have been explored [9][10], demonstrating low phase noise and 13.5 mW power consumption but a very limited tuning range [10]. Alternatively, high $Q$ overtone acoustic resonators leveraging the equally-spaced harmonics of a resonant cavity, such as high overtone bulk acoustic resonators (HBARs) [11] and lateral overtone bulk acoustic resonators (LOBARs) based on AlN [12], and AlN-on-SiC [13], were proposed. Unfortunately, their low $k_t^2$ and low figure-of-merit (FoM = $Q \cdot k_t^2$) in overmoded structures cannot enable a VCMO with a large tuning range and low power.

Recently, LOBARs based on lithium niobate (LiNbO$_3$) [14-18] have been shown with multiple high FoM resonances surpassing the state-of-the-art (SOA). Provided with an optimized oscillator design, they can potentially result in lower phase noise, lower power, and wider tuning range RF synthesizers suited for multi-mode IoT nodes.

This paper presents the first LiNbO$_3$ LOBAR-based VCMO. The VCMO is capable of locking to any of the ten overtones with a non-contiguous tunable oscillation frequency ranging from 300 to 500 MHz. It also shows a measured close-in phase noise of -100 dBc/Hz at 1 kHz offset from a 300 MHz carrier and a noise floor of -153 dBc/Hz while consuming 9 mW. The proposed VCMO can be used as a general-purpose oscillator [19] that applies to a variety of MEMS/crystal resonators in RF synthesizers.

## II. VCMO DESIGN

### A. LiNbO$_3$ LOBAR

The LOBAR consists of a suspended LiNbO$_3$ thin film and aluminum interdigitated electrodes (IDEs) that partially cover the top surface. The IDEs are alternately connected to the input and ground introducing a time-varying electric field which subsequently excites strain and stress standing waves inside the resonator cavity. The orientation of the device is chosen as −10° to +Y-axis in the X-cut plane of LiNbO$_3$ for exciting a family of shear horizontal (SH) modes of various lateral mode orders with high $k_t^2$.

In contrast to a conventional resonator targeting a single resonance [20-23], our LOBAR [15] features several equally-spaced resonances ranging from 100 to 800 MHz as shown in Fig. 1 (a). The center frequency ($f_c$) of the resonator and the frequency spacing ($\Delta_f$) between the adjacent tones can be set by the IDE pitch and resonator width, respectively. For resonances further away from $f_c$, the modes are less effectively excited, hence a reduced $k_t^2$ [15]. The VCMO is thus designed with the capability to lock to a maximum number of LOBAR overtones centered about $f_c$ (~ 400 MHz). As each overtone is characterized by a high FoM, oscillation at each overtone frequency only requires very low power to sustain. The main specs ($f_m$, $Q$, and $R_m$) of the resonator for the ten overtones spanning from 300 to 500 MHz are given in Table I. As a Zoomed-in example, the measured admittance of the 415 MHz resonance is shown in Fig. 1 (b), while the optical image of the device is shown in Fig. 1 (c).

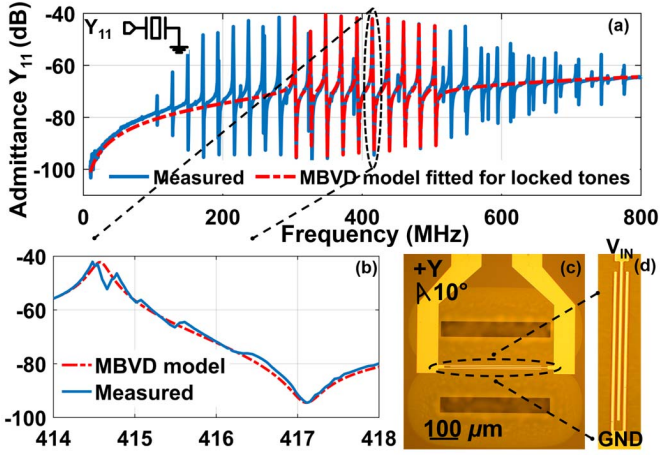

Fig. 1: (a) Measured and MBVD fitted responses of the LOBAR. (b) Measured admittance of the 415 MHz tone. (c) Optical image of the LOBAR. (d) Zoom-in picture of the LOBAR.

TABLE I.  $Q$ AND $R_M$ FOR THE TEN OVERTONES IN 300 – 500 MHZ RANGE

| Spec/Tones | 1 | 2 | 3 | 4 | 5 | 6 | 7 | 8 | 9 | 10 |
|---|---|---|---|---|---|---|---|---|---|---|
| $f_m$ (MHz) | 305 | 325 | 345 | 370 | 390 | 415 | 435 | 460 | 480 | 505 |
| $Q$ | 1650 | 1671 | 1945 | 1825 | 1908 | 1970 | 2608 | 2050 | 2202 | 3000 |
| $R_m$ | 122 | 225 | 107 | 115 | 125 | 130 | 127 | 147 | 167 | 175 |

### B. Reconfigurable Oscillator

The oscillator shown in Fig. 2 (a) consists of a LOBAR in a closed loop with two common emitter degenerated amplifiers and a voltage tunable varactor-embedded LC resonator. A common collector buffer is used to match the output to 50 Ω needed for the measurements. The tunable LC resonator is comprised of an inductor and varactor in parallel and loaded by two shunt capacitors ($C_S$) to the ground. By tuning the bias voltage $V_{VAR}$ of the varactor $C_P$, the LC tank adjusts the loop so that the Barkhausen conditions can be satisfied for a particular resonant mode and only one oscillation frequency is produced at the output.

The effects of $C_S$ and $C_P$ on the tank transfer function are shown in Fig. 3. $C_S$ controls only the tank's series resonant frequency ($f_s^t$) while $C_P$ controls both the series and parallel resonances ($f_p^t$). The smaller the $C_S$, the higher the $f_s^t$ and the smaller the tank's inductive range, i.e., bandwidth (BW= $f_p^t$ - $f_s^t$), hence affecting fewer overtones. On the other hand, the smaller the $C_P$, the larger the spacing between both resonant frequencies and BW, thus affecting more overtones.

Following the same framework, the spectrum can be divided into four regions with respect to $f_s^t$ and $f_p^t$ as shown in Fig. 4 (a). In Region 1 where frequencies are lower than $f_s^t$, the overtones fulfill the gain condition of oscillation but do not fulfill the phase condition. In Region 2 where the frequency aligns with $f_S$, the tone is coupled with a maximum gain. Furthermore, in Region 3 where frequencies lie within the tank BW closer to $f_p^t$, the tones get suppressed in gain despite fulfilling the phase condition. Finally, in Region 4 where frequencies are larger than $f_p^t$, the overtones do not fulfill both the gain and phase conditions. Therefore, the targeted LOBAR tone should be as close as possible to $f_s^t$ inside the tank inductive region. Fig. 4 (b) and (c) show the simulated loop gain and phase response of the VCMO for three different varactor bias voltages spanning the tuning range.

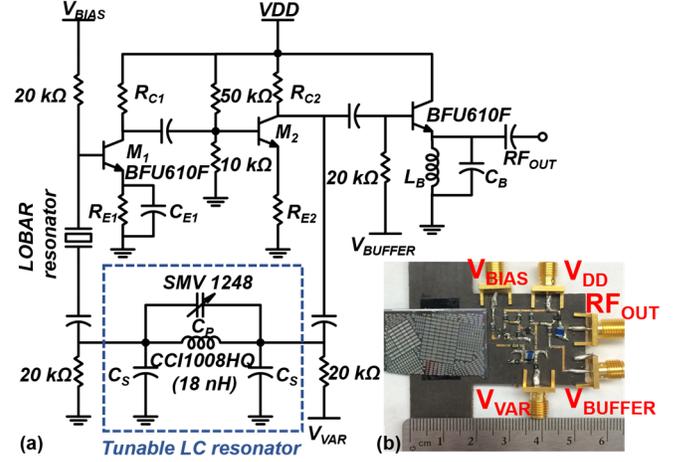

Fig. 2: (a) VCMO circuit schematic. (b) Board implementation.

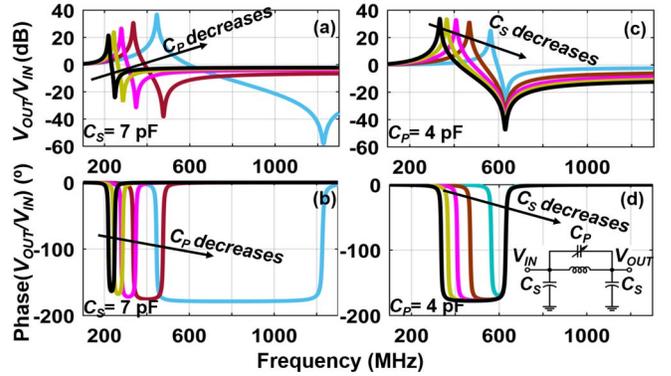

Fig. 3: (a) Gain and (b) phase responses with $C_S$ fixed to 7 pF and $C_P$ varying from 1 to 25 pF. (c) Gain and (d) phase responses with $C_P$ fixed to 4 pF and $C_S$ varying from 1 to 10 pF.

A silicon germanium BFU610F is chosen to implement the three transistors for its low noise figure and low power consumption. A CCI1008HQ of 18 nH is chosen for the tank inductor. This value allows locking to the modes with the lowest $R_m$ in the range of 300 to 500 MHz. An SMV1248 varactor is chosen to implement $C_P$ which can be varied from 22.62 pF to 1.3 pF as $V_{VAR}$ varies from 0 to 8 V.

$V_{BE}$ bias voltages of $M_1$ and $M_2$ are chosen based on three factors: First, a small BJT base current ($I_B$) value that gives a low flicker noise; Second, a reasonable gain to excite a single resonance without satisfying Barkhausen condition for more than one tone especially when the tones are very close with a $\Delta_f$ of 20 MHz; Third, low power consumption. The buffer design is borrowed from the millimeter-wave regime, where VCO buffers use a quarter wavelength stub to cancel the imaginary output impedance and match the real output impedance ($1/g_m$) of the buffer emitter to 50 Ω. For lower frequency designs, an LC tank can be used to mimic the stub. The tank reduces the power consumption in the buffer for the same output power

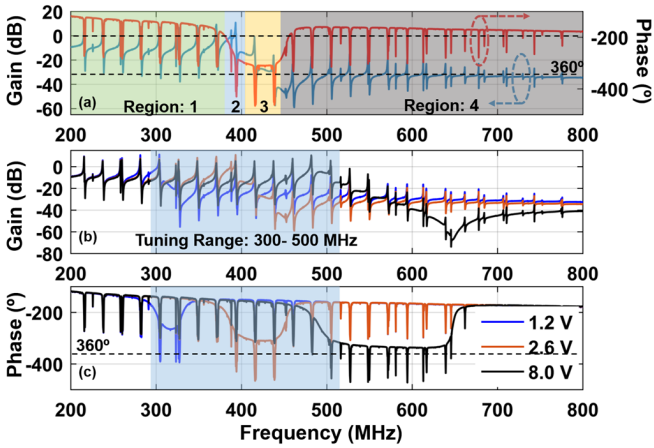

Fig. 4: (a) Illustration of four regions. (b) Simulated loop gain and (c) loop phase response for three different varactor bias voltages spanning the tuning range. Only three voltages are represented instead of ten for figure clarity.

when compared to a resistive loading.

## III. MEASUREMENTS

SMA connectors are used for all DC-biases to minimize any noise pick-up from the external sources. Fig. 2 (b) shows a PCB prototype with LiNbO$_3$ LOBAR sample bond-wired to the oscillator. A tunable oscillation frequency ranging from 300 to 500 MHz has been achieved by exploiting the ten overtones in the LOBAR. As shown in Fig. 5 (a), the VCMO shows continuous tuning near the series resonance of each overtone and a discrete hop of roughly 20 MHz when switching to an adjacent overtone. As shown in Fig. 5 (b), a specific V$_{VAR}$ can produce a maximum output power of 0 dBm for each tone across their continuous tuning range. The VCMO consumes only 9 mW in operation owing to the LOBAR's high FoM. Phase noise measurements were done with an Agilent E5052A Signal Source Analyzer and are reported in Fig. 6. The VCMO demonstrates on -100 dBc/Hz phase noise at 1 kHz offset from a 300 MHz carrier with a FoM$_{VCMO}$ of 200 dBc/Hz. The VCMO also achieves a best noise floor of -153 dBc/Hz due to high FoMs of LiNbO$_3$ LOBARs with a FoM$_{VCMO}$ of 193 dBc/Hz.

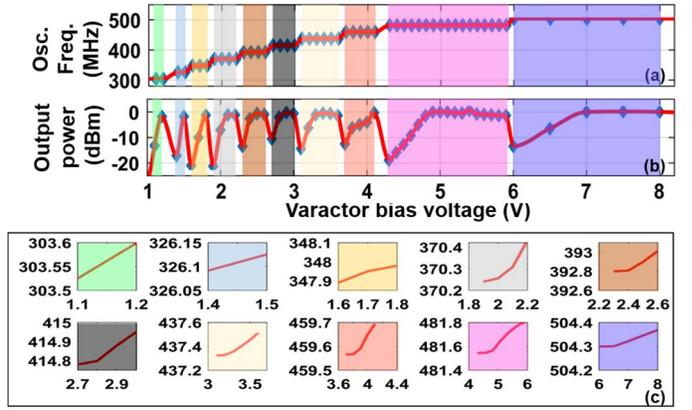

Fig. 5: (a) Oscillation frequency and (b) output power versus varactor bias. (c) Tuning response for each mode. Color codes are included for easy correlation.

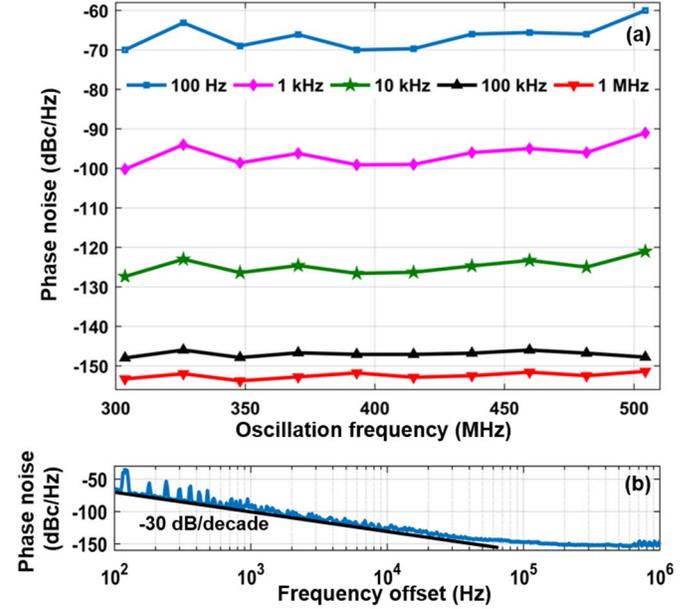

Fig. 6: (a) Phase noise of the 10 locked modes at different frequency offsets. (b) Phase noise plot of the 415 MHz mode.

TABLE II. PERFORMANCE SUMMARY AND COMPARISON TO RECONFIGURABLE MEMS OSCILLATORS

| Reference | | This work | [6] | | | | [7] | | | | [9] | | [10] | | [19] | | |
|---|---|---|---|---|---|---|---|---|---|---|---|---|---|---|---|---|---|
| MEMS | | LiNbO$_3$ LOBAR | AlN CMR | | | | AlN CMR | | | | AlN CMR | | AlN-on-Si | | SAW | | |
| Process | | Discrete | 0.5 μm CMOS | | | | 0.5 μm CMOS | | | | 0.5 μm CMOS | | 0.5 μm CMOS | | 0.18 μm CMOS | | |
| General purpose | | Yes | Yes | | | | Yes | | | | No | | No | | Yes | | |
| Number of resonances | | 10 | 4 | | | | 4 | | | | 2 | | 2 | | 3 | | |
| Number of resonators | | 1 | 4 | | | | 4 | | | | 1 | | 1 | | 3 | | |
| Frequencies (MHz) | | 300 - 500 (ten equally-spaced tones) | 268 | 483 | 690 | 785 | 176 | 222 | 307 | 482 | 472 | 1940 | 35 | 175 | 315 | 433 | 500 |
| dc power (mW) | | 9 | As high as 35.5 | | | | 10 | | | | - | 20 | 3.8 | 13.5 | 11.3 | 8.1 | 6.8 |
| Output power (dBm) | | 0 | 1.1 | 0 | -2.7 | -6 | -4.7 | -4.8 | -6.7 | -13.6 | -6.5 | -16 | 12.1 | 3.8 | - | - | - |
| PN (dBc/Hz) | 1 kHz | -100 | -94 | -88 | -83 | -70 | -79 | -88 | -84 | -68 | -82 | -69 | -112 | -103 | - | - | - |
| | 1 MHz | -153 | -93 | -92 | -90 | -78 | -74 | -85 | -84 | -72 | -86 | -85 | -93 | -98 | -134 | -141 | -135 |
| FoM$_{VCMO}$ (dBc/Hz) | 1 kHz | 200 | 187 | 186 | 184 | 172 | 174 | 185 | 184 | 172 | - | 182 | 197 | 197 | - | - | - |
| | 1 MHz | 193 | - | - | - | - | - | 197 | - | - | - | 205.7 | 167 | 173.6 | 173.4 | 184.6 | 180.6 |

The values in the shaded cells are referenced to a 300 MHz carrier.

FoM$_{VCMO}$ = $-L(\Delta f) + 20\log(\frac{f_o}{\Delta f}) - 10\log(\frac{P_{DC}}{1\,mW})$.

FoM$_{VCMO}$: VCMO's figure of merit, $f_o$: carrier frequency, $\Delta f$: frequency offset, $P_{DC}$: power consumed, $L(\Delta f)$: phase noise measured at an offset $\Delta f$ from the carrier.

Fig. 6 (b) shows the phase noise profile for the 415 MHz carrier as an example. The spurious profile is believed to be a result of the spurious resonance mode shown in Fig 1 (b).

IV. CONCLUSIONS

Table II shows a performance summary and comparison to other reconfigurable MEMS oscillators. This paper presents the highest number of locked tones of a single acoustic resonator with competitive phase noise and $FOM_{VCMO}$ results, making $LiNbO_3$ LOBAR VCMO a great candidate for direct RF synthesis deployed in wireless transceivers targeting multi-mode IoT applications. The tuning range and power consumption can be further enhanced via implementing the active circuitry in a recent-node CMOS. Moreover, having a switchable bank of tunable LC tanks with different inductor values would allow the VCMO to harness all the overtones provided by the LOBAR for a broader tuning range. This VCMO can also be scaled up in frequency for 5G by coupling with four mm-wave overtone resonances provided by a recently advanced A1 $LiNbO_3$ resonator [24].